\documentclass[nofootinbib,prd,aps,twocolumn,preprintnumbers,amsmath,amssymb,superscriptaddress]{revtex4}
\usepackage{amsmath}
\usepackage{amssymb}
\usepackage{graphicx}
\usepackage{subfigure}
\usepackage{color}
\usepackage[colorlinks,linkcolor=magenta,anchorcolor=blue,citecolor=green]{hyperref}
\usepackage{ulem}
\usepackage{pifont}
\usepackage{makecell}
\begin{document}

\title{Perturbations of Mimetic Curvaton}

\author{Anxianyi Xiong}
\email{u202010150@hust.edu.cn}
\affiliation{School of Physics, Huazhong University of Science and Technology\\
Wuhan, 430074, China}

\author{Xin-zhe Zhang}
\email{zincz@hust.edu.cn}
\affiliation{School of Physics, Huazhong University of Science and Technology\\
Wuhan, 430074, China}

\author{Taotao Qiu}
\email{qiutt@hust.edu.cn(corresponding author)}
\affiliation{School of Physics, Huazhong University of Science and Technology\\
Wuhan, 430074, China}

\begin{abstract}
The mimetic gravity theory is one of the interesting modified gravity theories, which aims to unify the matter component of our universe within the power of gravity. The mimetic-like theory can also be responsible for primordial perturbations production, e.g., when the mimetic field is set to be like a curvaton field, and the adiabatic perturbation can thus be generated from the isocurvature perturbation via usual curvaton mechanism \cite{Zhang:2022bde}. In the original mimetic curvaton model, the parameter $\lambda$ was purely an algebraic multiplier, lack of any perturbed dynamics. In the current paper, we treat $\lambda$ as an auxiliary field, with its perturbation $\delta\lambda$ evolving alongside. We show that, with such a consideration, the adiabatic perturbation can still be generated from the curvaton mechanism, and becomes scale invariant with different field space configurations.
\end{abstract}

\maketitle

\section{Introduction}

It had become widely recognized that the gravity theory which controls our universe is likely to be Einstein's General Relativity (GR). However, it is proved that the pure GR has nothing other than 2 tensor degrees of freedom (DOF), which would no longer get along with the development of modern physics in recent years, such as the appearance of evidences of existence of dark matter, dark energy and acceleration in the early stages of the universe. It is realized that to properly describe the universe, new DOFs are inevitable. 

Normally, one can add new DOFs by directly introducing them into the gravity theory, such as new fields or new matters. Recently, people find a new way of enhancing DOFs, relating to the disformal transformations \cite{Bekenstein:1992pj}. Usual disformal transformations, as is well known, do not alter the number of DOFs, except the cases where they are singular and not invertible \cite{Deruelle:2014zza, Arroja:2015wpa, Domenech:2015tca, Domenech:2023ryc}. If one perform the transformation as
\begin{align}
    g _{\mu \nu} \rightarrow \tilde{ g} _{\mu \nu} ~, ~~~ \text{where} ~~~ g _{\mu \nu} =(\tilde{ g} ^{\alpha \beta} \partial _{\alpha} \phi \partial _{\beta} \phi) \tilde{ g} _{\mu \nu} ~,
    \label{disformal transformation}
\end{align}
where $g _{\mu \nu}$ is the metric, $\tilde{ g} _{\mu \nu}$ is auxiliary metric and $\phi$ is a scalar field. One could find one more scalar DOF (longitudinal DOF) proportional to the trace of $(G ^{\mu} _{\nu} - T ^{\mu} _{\nu})$. This disformal transformation generates a new gravity theory different from the original GR, which was introduced first by Chamseddine and Mukhanov, dubbed as ``mimetic gravity" \cite{Chamseddine:2013kea}. In the original mimetic gravity theory, the additional DOF can be regarded as an effective part of energy density with vanishing pressure, therefore performing the effect as dark matter \cite{Chamseddine:2013kea, Golovnev:2013jxa}. Since Eq. \eqref{disformal transformation} is equal to the equality $g ^{\mu \nu} \partial _{\mu} \phi \partial _{\nu} \phi =1$, this can also be extended to the form with a Lagrangian multiplier term $\lambda (g ^{\mu \nu} \partial_\mu\phi\partial_\nu\phi-1)$ added to the gravity action, where $\phi$ is the auxiliary field. Such a theory can be applied to various aspects of cosmology, see \cite{Sebastiani:2016ras} for the comprehensive review. 

Nonetheless, people find that if there is only one DOF added, it is not propagating and act as a merely constrained DOF \cite{Lim:2010yk}, so does not work at the perturbation level when applied to the inflation \cite{Chamseddine:2014vna}. This problem can be solved by simply adding more DOFs. One way of doing this is to have higher derivative terms of the auxiliary field, such as $\Box \phi$ \cite{Chamseddine:2014vna}, but leading to ghost or gradient instabilities with a pure $\Box \phi$ term \cite{Firouzjahi:2017txv, Ijjas:2016pad, Takahashi:2017pje}. There are many discussions about this issue, seen in \cite{Mirzagholi:2014ifa, Chaichian:2014qba, Hirano:2017zox, Zheng:2017qfs, Gorji:2017cai, Casalino:2018tcd, Casalino:2018wnc, HosseiniMansoori:2020mxj}. Another way is to add more auxiliary fields \cite{Firouzjahi:2018xob, Shen:2019nyp, Mansoori:2021fjd, Zheng:2022vwm}. If all the auxiliary fields are canonical, there would be no ghost or gradient instabilities but the non-propagating adiabatic mode, as \cite{Mansoori:2021fjd} has noticed, which will thus cause the problem. Note that it is also interesting to consider modified mimetic gravity, see \cite{Nojiri:2014zqa} for an example. 

In order to tackle such a non-propagating issue of the multi-field mimetic theory, in recent works \cite{Zhang:2022bde} the authors resorted to the curvaton mechanism \cite{Lyth:2001nq, Lyth:2002my}. They suggested that, although the adiabatic mode does not propagate during inflationary epoch, it can be sourced by the isocurvature mode which can propagate, and after inflation we can get the final adiabatic perturbations. In \cite{Zhang:2022bde}, it is assumed for simplicity that the Lagrangian multiplier $\lambda$ is just a algebraic multiplier, and therefore its perturbation $\delta \lambda$ has been neglected. In this paper, however, we're going to consider a more complete case, by treating the multiplier to be an non-dynamical field, therefore the perturbation $\delta \lambda$ appears as well. Note that the newly introduced perturbation variable does not mean the increase of the total number of DOFs, for which we will have one more constraint equation for $\delta \lambda$ as well, but it will get involved to the perturbation equations, and may change the evolution behaviors of the perturbations. It is therefore interesting to see that in this case, whether the curvaton mechanism still works or not.

The consequent contents are organized as follows: in Sec. \ref{mimetic review} we briefly review the multi-field mimetic inflation model in general, as well as the mimetic curvaton model introduced in \cite{Zhang:2022bde}. The next two sections are the main part of our current work, where we discuss the perturbations of mimetic curvaton with the presence of $\delta \lambda$ and a different field-space configuration. In Sec. \ref{infepoch} we consider the inflationary epoch, while in Sec. \ref{MDepoch} we consider the matter-dominated epoch. In both epochs, the results of background evolution and perturbations are presented. Finally, Sec. \ref{conclusions} includes the conclusion and final discussions.

%Since the observation of flat rotation curve in galaxies, we have confirmed that there should be some new physics beyond what we have know...

\section{Mimetic Inflation and Curvaton Models}
\label{mimetic review}
\subsection{background}
The generalized multi-field mimetic inflation model will have the action form as the following \cite{Chamseddine:2014vna, Zhang:2022bde}:
\begin{align}
    S &= \int d^{4}x \sqrt{ -g} \bigg[ \frac{ M _{\text{P}} ^{2}}{ 2} R \nonumber\\
    & -\frac{ 1}{ 2} \lambda \left( G _{a b} g ^{\mu \nu} \partial _{\mu} \phi ^{a} \partial _{\nu} \phi ^{b} +\Omega ^{2} \right) -V( \phi ^{a}, \lambda) \bigg] ,
    \label{action of mimetic gravity}
\end{align}
where $\lambda$ is a Lagrangian multiplier, $G _{a b} \equiv G _{a b} (\phi ^{a})$ is the metric of the field space, and $\Omega$ is a constant corresponding to the typical energy scale of $\phi ^{a}$. Since now we consider $\lambda$ to be an auxiliary field as well, we extend the potential to contain also $\lambda$. In FLRW metric one can obtain the Friedmann equations,
%\begin{align}
 %   \begin{cases}
  %      3 M _P ^2 H ^2 = V(\phi^a) + \lambda \dot \phi _a \dot \phi ^a \\
   %     - 2 M _p ^2 \dot{ H } = \lambda ( \dot{ \phi } ^a \dot{ \phi } _a + \partial _i \phi ^a \partial ^i \phi _a )
    %\end{cases} ,
    %\label{Friedmann equtions general}
%\end{align}
%where $H \equiv \dot{ a } / a$ is Hubble parameter and $\partial _i \phi ^a \partial ^i \phi _a \equiv a ^{ -2} \delta ^{ i j} \partial _i \phi ^a \partial _j \phi _a$. From the first principle of cosmology, matter fields could be isotropic and homogeneous. Thus the Friedmann equations \eqref{Friedmann equtions general} becomes
\begin{eqnarray}
    \label{Friedmann equations FLRW}
    3 M _{\text{P}} ^{2} H ^{2} &=& V +\lambda \left( \frac{ 1}{ 2} \dot \phi _{a} \dot \phi ^{a} +\frac{ 1}{ 2}\Omega ^{2} \right) ~, \text{and}\\
    2 M _{\text{P}} ^{2} \dot{ H} &=& -\lambda (\dot \phi _{a} \dot \phi ^{a}) ~,
\end{eqnarray}
where $H \equiv \dot{ a} /a$ is Hubble parameter, and $\dot \phi _{a} \dot \phi ^{a} \equiv G _{a b} \dot \phi ^{a} \dot \phi ^{b}$. Varying the action with respect to $\lambda$, one can get the constraint equation:
\begin{align}
    \frac{ 1}{ 2} \dot \phi _{a} \dot \phi ^{a} -\frac{ 1}{ 2}\Omega ^{2} -V _{\lambda} =0 ~,
    \label{constraint}
\end{align}
where $V _{\lambda} \equiv dV /d\lambda$. This makes the Friedmann equations \eqref{Friedmann equations FLRW} become:
\begin{eqnarray}
        3 M _{\text{P}} ^{2} H ^{2} &=& V +\lambda \left( V _{\lambda} + \Omega ^{2} \right) ~, \text{and}\\
        2 M _{\text{P}} ^{2} \dot{ H} &=& - \lambda (2V _{\lambda} +\Omega ^{2}) ~ .
    \label{Friedmann equations FLRW2}
\end{eqnarray}
Note that when the potential does not depend on $\lambda$, the Friedmann equations reduce to the normal one in \cite{Chamseddine:2014vna, Zhang:2022bde}. It is also straightforward to get the equations of motion for the mimetic fields $\phi ^{a}$:
\begin{align}
    \Tilde{ D} _{t} \dot \phi ^{a} + \left( 3 H +\frac{ \dot{ \lambda}}{ \lambda} \right) \dot \phi ^{a} +\frac{ 1}{ \lambda} V ^{ ,a} = 0 ,
    \label{eom of fields}
\end{align}
where $\Tilde{ D} _{t} X ^{a} \equiv \dot{ X} ^{a} + \Tilde{ \Gamma} ^{a} _{b c} \dot{ \phi} ^{b} X ^{c}$ with $\Tilde{ \Gamma} ^{a} _{b c}$ is Christoffel connection of the field space and $V ^{, a} \equiv G ^{a b} \partial V( \phi ^{c}) /\partial \phi ^{b}$. 

The continuous equation gives the evolution of Lagrangian multiplier $\lambda$, which is
\begin{align}
(\Omega ^{2} +V _{\lambda}) \dot{ \lambda} +6 H \lambda( \frac{ 1}{ 2}\Omega ^{2} +V _{\lambda}) +\lambda \dot{ V} _{\lambda} +\dot{ V} =0 ~ .
    \label{eom of lambda}
\end{align}
Due to the involvement of $V_\lambda$, the solution of the above equation is not so obvious. However, we can still get some ansatz solutions by requiring that the system drives an inflationary period, with a matter-dominant epoch following. In the inflationary period where generally the slow-roll condition must be satisfied, it is natural to require a nearly flat potential, and $V \gg \lambda$. In this case, one has $V _{\lambda} \simeq 0$. On the other hand, in the matter-dominant epoch, one requirement of the solution is to have $V\ll \lambda\propto a^{-3}$, while $V_\lambda$ is nearly a constant. For the first condition, $\lambda \propto a ^{-3}$, Eq. \eqref{eom of lambda} is derived as:
\begin{align}
    \lambda \dot{V}_\lambda+V_{\phi^a}\dot\phi^a=0~,
    \label{conditionmd}
\end{align}
with $\dot{ V} =V _{\lambda} \dot{ \lambda} +V _{\phi ^{a}} \dot{ \phi} ^{a}$ and $\dot{ \lambda} = -3 H \lambda$. For the second condition, $V_\lambda$ is a constant, we can require that the potential linearly depends on $\lambda$, which further gives constraints on the dependence of the potential on the fields: $V _{\phi ^{a}} \dot{ \phi} ^{a} =0$. This new constraint should be taken in our following model building.

%and the solution of $\lambda$ is
%\begin{align}
%    \lambda = - \frac{ 1 }{ 2 } a ^{ - 3 } ( \int dt ^\prime a ^3 \dot{ V } + C _\lambda) ,
%    \label{sol of lambda}
%\end{align}
%where $C _\lambda$ is a constant. In matter dominated universe, we have $\dot{ V }$ vanished, and $\lambda \propto a ^{ - 3 }$ gives the component of dark matter. Thus, a intriguing possibility is that the quantum potential $V( \phi ^a )$ is not vanish at the early universe, and then decreases with the decay of matter fields turning the vacuum energy into the dark matter. The beginning of this process is solely the inflation introduced by \cite{Zhang:2022bde}.
\subsection{perturbations}
Next we study the cosmological perturbations of the system. We perturb the mimetic fields as
\begin{align}
    \phi ^{a} = \phi ^{a}( t) +\delta \phi ^{a}( t, x) ~, ~ \lambda = \lambda _{0}( t) +\delta \lambda( t, x) ~,
\end{align}
and the metric in the co-moving spatial flat gauge,
\begin{align}
    ds ^{2} =( -N ^{2} +N _{i} N ^{i}) dt ^{2} +2 N _{i} dt dx ^{i} +h _{i j} dx ^{i} dx ^{j} ,
    \label{ADM metric}
\end{align}
where $N = 1 + n$ is the lapse function, $N _{i} =\partial _{i} B$ is the shift function and $h _{i j}$ is the metric of the spatial hyper-surface. The second order perturbed action is:
\begin{align}
    S ^{(2)} &=\int d^{4}x \bigg( -\frac{ 1}{ 2} a ^{3} \lambda \partial _{\mu} \delta \phi _{a} \partial ^{\mu} \delta \phi ^{a} -\frac{ 1}{ 2} \mathcal{ M} ^{2} _{\ a b} \delta \phi ^{a} \delta \phi ^{b} \nonumber \\
    &+a ^{3} \delta \lambda \bigg( \dot \phi _{a} \delta \dot \phi ^{a} -\frac{ 1}{ 2} \epsilon H \delta \phi _{a} \dot \phi ^{a} +\frac{ 1}{ 2} G _{b c ,a} \dot \phi ^{b} \dot \phi ^{c} \delta \phi ^{a} \nonumber\\
    &+\frac{ 1}{ 2} V _{ ,a \lambda} \delta \phi ^{a} \bigg) \bigg) ,
    \label{action of perturbations}
\end{align}
where the mass matrix
\begin{align}
    \mathcal{ M} ^{2} _{\ a b} &=a ^{3} \bigg( \frac{ 1}{ 2} V \epsilon ^{2} H ^{2} G _{a c} G _{b d} \dot \phi ^{c} \dot \phi ^{d} -\frac{ 1}{ 2} \lambda G _{c d ,a b} \dot \phi ^{c} \dot \phi ^{d} \nonumber \\
    &+\frac{ 1}{ 2} \lambda \epsilon H G _{c d ,a} G _{b e} \dot \phi ^{c} \dot \phi ^{d} \dot \phi ^{e} + \epsilon H V _{, a} G _{b c} \dot \phi ^{c} +V _{, a b} \bigg) \nonumber \\
    &-\frac{ d}{ d t} \bigg( \frac{ 1}{ 2} a ^{3} \lambda \epsilon H G _{a c} G _{b d} \dot \phi ^{c} \dot \phi ^{d} -a ^{3} \lambda G _{b c ,a} \dot \phi ^{c} \bigg) ~,
    \label{mass matrix of perturbations}
\end{align}
and $\epsilon \equiv -\dot H /H ^{2}$ is the slow-roll paramter. Variation with respect to $\delta \lambda$ gives a constraint on perturbations:
\begin{align}
    \mathcal{ Z} _{\delta \lambda} &=\dot \phi _{a} \delta \dot \phi ^{a} -\frac{ 1}{ 2} \epsilon H \dot \phi _{a} \delta \phi ^{a} +\frac{ 1}{ 2} G _{b c ,a} \dot \phi ^{b} \dot \phi ^{c} \delta \phi ^{a} +\frac{ 1}{ 2} V_{, a \lambda} \delta \phi ^{a} \nonumber \\
    &=0 ,
    \label{mimetic constrain of perturbations}
\end{align}
which is the perturbed mimetic constraint \eqref{constraint}. The entire Hamiltonian analysis can been seen in \cite{Mansoori:2021fjd}. 

The equations of motion for the field-perturbations are:
\begin{align}
    &\delta \ddot \phi ^{a} -\partial _{i} \partial ^{i} \delta \phi ^{a} + \bigg( 3 H +\frac{ \dot \lambda}{ \lambda} \bigg) \delta \dot \phi ^{a} + G ^{a} _{\ b ,c} \dot \phi ^{c} \delta \dot \phi ^{b} \nonumber \\
    &+\frac{ 1}{ a ^{3} \lambda} \mathcal{ M} ^{2 a} _{\ \ b} \delta \phi ^{b} +\dot \phi ^{a} \frac{ \delta \dot \lambda}{ \lambda} \nonumber \\
    &+\bigg( \frac{ 1}{ 2} \epsilon H \dot \phi ^{a} -\frac{ 1}{ 2} V ^{, a} _{\lambda} -\frac{ 1}{ \lambda} V ^{, a} -\frac{ \dot \lambda}{ \lambda} \dot \phi ^{a} \bigg) \frac{ \delta \lambda}{ \lambda} =0 ~.
    \label{eom of perturbations of fields}
\end{align}
Substitute Eq. \eqref{eom of perturbations of fields} into the time derivative of the perturbed constraint \eqref{mimetic constrain of perturbations}, one can obtain the equation of motion of $\delta \lambda$ as:
\begin{align}
    &\dot \phi _{a} \dot \phi ^{a} \frac{ \delta \dot \lambda}{ \lambda} +\bigg( \frac{ 1}{ 2} \epsilon H \dot \phi _{a} \dot \phi ^{a} -\frac{ 1}{ 2} V _{, a \lambda} \dot \phi ^{a} -\frac{ 1}{ \lambda} V _{a} \dot \phi ^{a} -\frac{ \dot \lambda}{ \lambda} \dot \phi _{a} \dot \phi ^{a} \bigg) \frac{ \delta \lambda}{ \lambda} \nonumber \\
    &+\dot \phi _{a} \delta \ddot \phi ^{a} -\dot{ \mathcal{Z}} _{\delta \lambda} -\dot \phi _{a} \partial _{i} \partial ^{i} \delta \phi ^{a} +\bigg( 3 H +\frac{ \dot \lambda}{ \lambda} \bigg) \dot \phi _{a} \delta \dot \phi ^{a} \nonumber \\
    &+G _{a b ,c} \dot \phi ^{a} \dot \phi ^{c} \delta \dot \phi ^{b} +\frac{ 1}{ a ^{3} \lambda} \mathcal{ M} ^{2} _{\ a b} \dot \phi ^{a} \delta \phi ^{b} =0 ~.
    \label{eom of delta lambda}
\end{align}

Due to the perturbed constraint equation, the perturbation of the multiplier $\delta\lambda$ is not a DOF, and the system has the same DOF as the usual multi-field system \cite{Shen:2019nyp, Mansoori:2021fjd}. However, as \cite{Mansoori:2021fjd} has found, at least in the case where $V$ is independent on $\lambda$, the perturbed mimetic constraint will combine the adiabatic and entropy modes in the following form,
\begin{align}
    \dot{ u } _T = \epsilon H u _T + \dot \Theta u _N
\end{align}
where $u _T$ and $u _N$ are the adiabatic perturbation and the entropy perturbation respectively and $\dot \Theta$ is the angular velocity of perturbations' trajectory in the field space. This will make the kinetic term of the adiabatic perturbation disappear in the perturbed action \eqref{action of perturbations}, causing the seemingly non-propagation of the adiabatic perturbation.

\subsection{curvaton}

In \cite{Zhang:2022bde}, we discuss the curvaton mechanism in the framework of mimetic gravity theory, where the adiabatic perturbation can be generated from the entropic one. We assume the first of the two fields, $\varphi$, to be the inflaton field, while the second one, $\theta$, to be the curvaton field. Technically, we chose a very specific form of the field metric $G _{a b}$:
\begin{align}
    G _{a b} =\text{diag} \left \lbrace 1 ,6 \sinh ^{2} \left( \frac{ \varphi}{ \sqrt{ 6} M} \right) \right \rbrace ~ ,
\end{align}
where $M$ is a constant respect with the typical scale of $\varphi$. Moreover, for simplicity we consider the parameter $\lambda$ to be an algebraic multiplier, therefore the perturbation of $\lambda$ was turned off. For the potential form, we choose $V \approx \tanh ^{2 n}( \varphi /\sqrt{ 6} M)$ for inflationary epoch. Such kind of potential satisfies the slow-roll condition which is required by inflation, and gives rise to a solution of the two fields with constant-velocities: $\dot \varphi \simeq \pm \sqrt{ 2} M ^{2}$, $\dot \theta \simeq 0$.  In the matter-dominated epoch when the potential reaches its minimum, we set its form to be $V \simeq \lambda[ m _{\text{eff}} ^{2} \varphi ^{2} +\alpha \ln( \varphi /\varphi _{1})]$, where the $\alpha$-term is a small correction to the main quadratic form and $\varphi_1$ is a constant. This will induce a solution of a semi-oscillating $\varphi$: $\varphi \simeq \sqrt{ \sin( m _{\text{eff}} t)}$, and $\dot \theta \simeq( m _{\text{eff}} t) ^{-1}$. Moreover, for the curvaton model, we turned off all the metric perturbations. Making use of the perturbation equations \eqref{eom of perturbations of fields}, one gets the solutions for $\delta\varphi$ and $\delta\theta$, respectively,
\begin{align}
    \delta \varphi \simeq \delta \theta \simeq \frac{ H}{ 2 \pi} ~,~~~\text{for inflationary epoch}, \text{ and}\\
    \begin{cases}
        \delta \varphi &=\frac{ 1}{ \varphi} \big( D _{+} e ^{i k \tau} +D _{-} e ^{-i k \tau} \big), \\
        \delta \theta &=\frac{ \sqrt{ 3}}{ 3 A m _{\text{eff}} t} \big( D _{+} e ^{i k \tau} +D _{-} e ^{-i k \tau} \big),
    \end{cases} \text{for MD epoch}.
\end{align}
When the curvaton dominates the universe or decays into the background, its perturbation will be transferred into the curvature perturbation, $\zeta \equiv -H \delta \rho /\dot \rho$. In detail, $\zeta$ will be dependent on both $\delta \varphi$ and $\delta \theta$. According to the calculations in \cite{Zhang:2022bde}, the $\delta \varphi$-containing term will dilute as time goes, therefore $\zeta$ will only depend on $\delta \theta$, in the form of mean square root. After some manipulations, we finally get the power spectrum of the curvature perturbation as:
\begin{align}
    P _{\zeta} \equiv \frac{ k ^{3}}{ 2 \phi ^{2}} |\zeta| ^{2} \sim r ^{2} \frac{ H ^{2}}{ 2 \pi ^{2} M _{p} ^{2} \epsilon \dot \theta _{i} ^{2}} ~ ,
\end{align}
where $r \equiv \rho _{\theta} /\rho _{\text{tot}}$ is the energy fraction of $\theta$, and $\dot \theta _{i}$ is the value of $\dot \theta$ at the beginning of the matter-dominated era. 

\section{inflation epoch}
\label{infepoch}
\subsection{background}
In this section, first of all we give a specific case of the mimetic curvaton model, which can drive a inflationary period. For simplicity, we re-scale the  choose the field-space metric to be the trivial one:
\begin{align}
    G _{a b} =\text{diag} \lbrace 1 ,1 \rbrace ~ .
    \label{field metric}
\end{align}
According to the discussions in the last section, we need to choose a flat potential in order to satisfy the slow-roll condition. One of such a potential is
\begin{align}
    V( \varphi ,\theta ,\lambda) =V( \varphi) =3 \tanh ^{2} \left( \frac{ \varphi}{ \sqrt{ 6 \Omega}} \right) ~ .
\label{potentialinf}
\end{align}
Therefore, according to the field equations \eqref{eom of fields}, one has
\begin{align}
    \ddot \varphi +\big( \Omega ^{2} -\dot \varphi ^{2} \big) \frac{ V _{\varphi}}{ \lambda \Omega ^{2}} =0 ~ ,\ 
    \ddot \theta -\frac{ V _{\varphi} \dot \varphi}{ \lambda \Omega ^{2}} \dot \theta =0 ~ .
    \label{eom of fields2}
\end{align}
These equations give the solution as
\begin{align}
    \varphi =\varphi _{0} -\Omega t ~ ,\ 
    \dot \theta =\dot \theta _{0} \exp \left( \int \frac{ V _{\varphi} \dot \varphi}{ \lambda \Omega ^{2}} dt \right) \approx 0 ~ .
    \label{solutions of fields in inflation}
\end{align}
On the other hand, the equation for $\lambda$ becomes:
\begin{align}
    \dot \lambda +3 H \lambda -\frac{ \sqrt{ 6}}{ \Omega^{3/2}} \cosh ^{-2} \left( \frac{ \varphi}{ \sqrt{ 6\Omega}} \right) \tanh \left( \frac{ \varphi}{ \sqrt{ 6\Omega}} \right) =0 ~ .
    \label{eom of lambda2}
\end{align}
which is difficult to obtain an exact solution directly. However, we can discuss about its approximate solution during different stages in inflation. From now on we take $\Omega =1$ to briefly obtain analytical solutions in this section. Since the potential \eqref{potentialinf} with the behavior as \eqref{solutions of fields in inflation} is a large-field kind of inflaton potential, it brings the $\varphi$ field from a large value to a small one. Therefore, at the early time of inflation when $\varphi \gg 1$, the last term in Eq. \eqref{eom of lambda2} can be neglected, giving rise to an approximate solution of $\lambda$:
\begin{align}
    \lambda =\lambda _{0} e ^{-3 H t} ~ .
\end{align}
Meanwhile, at the late time of evolution when $\varphi \lesssim 1$, the last term in Eq. \eqref{eom of lambda2} cannot be neglected. Since the inflation is still not ending, we have 
\begin{align}
    V _{\varphi} \sim \sqrt{ 6} e ^{-\sqrt{ 2 /3}( \varphi _{0} -t)} ~ .
\end{align}
The $\lambda$ has the solution as:
\begin{align}
    \lambda =\frac{ \sqrt{ 6}}{ \sqrt{ 2 /3} +3} e ^{\sqrt{ 2 /3}( t -\varphi _{0})} ~ .
\end{align}

\begin{figure}[h]
    \subfigure[]{
    \includegraphics[width=200.pt]{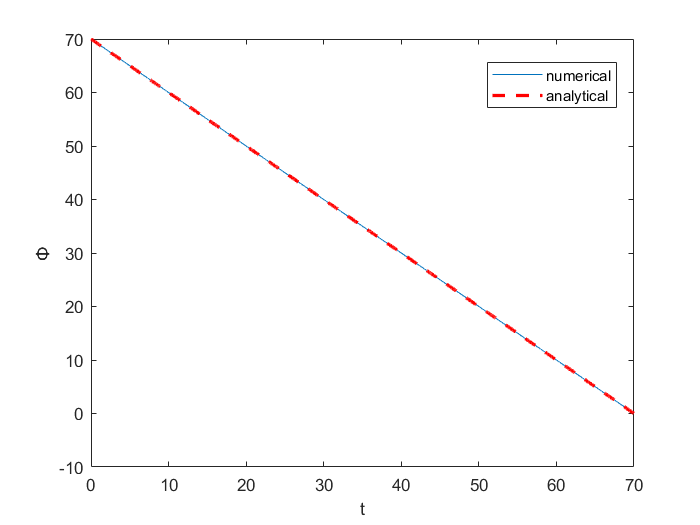}
    }
    \subfigure[]{
    \includegraphics[width=200.pt]{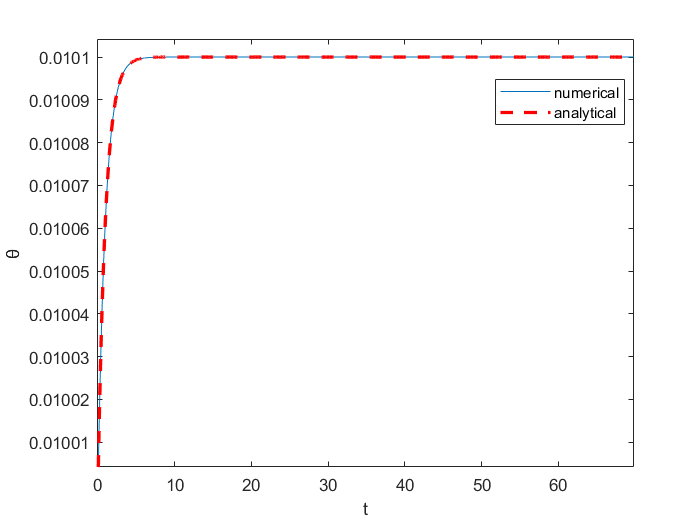}
   }
      \subfigure[]{
    \includegraphics[width=200.pt]{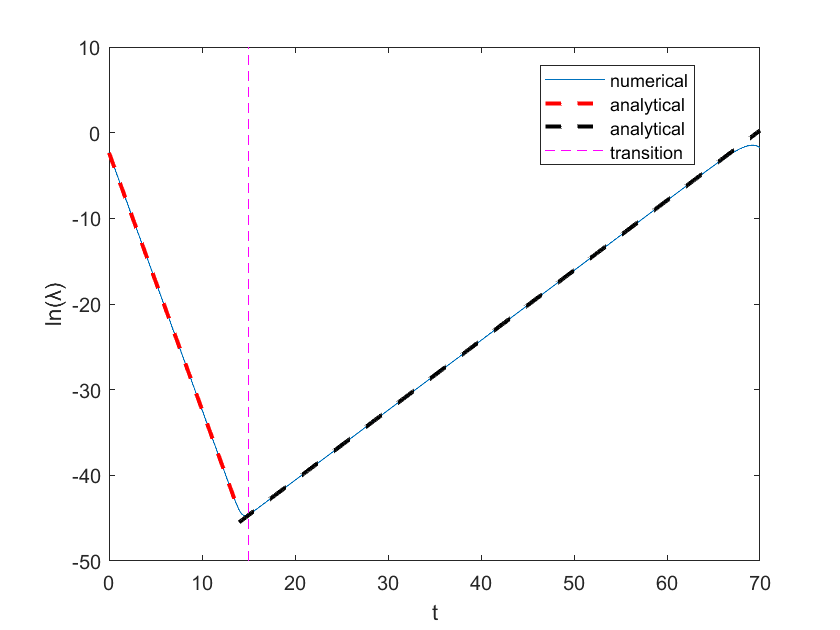}
    }
    \caption{Plots of $\varphi$, $\theta$ and $\ln(\lambda)$ with respect to time $t$. The initial values are chosen to be $\lambda _{0} =10 ^{-4}$, $\varphi _{0} =70 \Omega /M _{\text{P}}$, $\theta _{0} =10 ^{-2} \Omega /M _{\text{P}}$ and $\dot \theta _{0} =10 ^{-4} \Omega$. Dash lines show the analytical results and solid lines show the numerical results.}
    \label{fieldplotinf}
\end{figure}

In order to verify with the above analysis, We also perform numerical calculations for Eqs. \eqref{eom of fields2} and \eqref{eom of lambda2}, and plot the evolution of the fields $\varphi$, $\theta$ and $\lambda$ in Fig. \ref{fieldplotinf}. From the plot we can see that, the inflaton field $\varphi$ evolves with nearly a constant velocity, while $\theta$ is nearly constant with a vanishing velocity. The evolution of $\varphi$ and $\theta$ also satisfies the constraint equation \eqref{constraint}. Moreover, one can see that $\lambda$ will have a ``turn-around" behavior, and the turn-around time will be determined by the specific choice of initial values and the form of the potential.
\subsection{perturbations}

In this section we discuss about the perturbations generated during the inflationary period. For two-field system, %it is useful to apply the spatial-flat gauge where the metric perturbations is:
%\begin{align}
%    ds^2=-(1+2\Phi)dt^2+2\partial_iBdx^idt+a^2(t)\delta_{ij}dx^idx^j~,
%\end{align}
%here $\Phi$ is the Newtonian potential and $B$ is a scalar function. {\red 
applying the curvaton mechanism with no metric perturbation one can obtain the equations of field-perturbations as
\begin{eqnarray}
    \delta \ddot \phi ^{a} +(3 H +\frac{ \dot \lambda}{ \lambda}) \delta \dot \phi ^{a} +(\frac{ k ^{2}}{ a ^{2}} +\frac{ 1}{ \lambda} V _{\phi ^{a} \phi ^{a}}) \delta \phi ^{a} \nonumber \\
    -\frac{ 1}{ 2 \lambda} V _{\phi ^{a} \phi ^{b}} \delta \phi ^{b} +\frac{ \delta \lambda}{ \lambda}( \ddot \phi ^{a} +3 H \dot \phi ^{a} +V _{\phi ^{a} \lambda}) =0 ~ ,
\end{eqnarray}
and the perturbed constraint equation becomes
\begin{align}
    \dot \varphi \delta \dot \varphi +\dot \theta \delta \dot \theta -V _{\lambda \varphi} \delta \varphi -V _{\lambda \theta} \delta \theta -V _{\lambda \lambda} \delta \lambda =0 ~ .
\end{align}
In inflationary epoch, we have $V =V( \varphi)$, $\dot \lambda /H \lambda \ll 1$. Meanwhile, the background level of the fields $\varphi$ and $\theta$ satisfy Eq. \eqref{eom of fields2}. Then the above equations turn out to be
\begin{eqnarray}
    \delta \ddot \varphi +3 H \delta \dot \varphi +\bigg( \frac{ k ^{2}}{ a ^{2}} +\frac{ V _{\varphi \varphi}}{ \lambda} \bigg) \delta \varphi \nonumber \\
    +(\ddot \varphi +3 H \dot \varphi) \frac{ \delta \lambda}{ \lambda} +\dot \varphi \frac{ \delta \dot \lambda}{ \lambda} &=&0 ~ , \\
    \delta \ddot \theta +3 H \delta \dot \theta +\frac{ k ^{2}}{ a ^{2}} \delta \theta &=&0 ~ , \\
    \dot \varphi \delta \dot \varphi &=&0 ~ .
\end{eqnarray}
From the second and the third equation one can immediately get the solution for $\delta\varphi$ and $\delta\theta$ during inflation:
\begin{align}
    \delta \varphi \sim \text{const.} \times k ^{3 /2} ~,~~~\delta \theta \simeq \frac{ H}{ 2 \pi} ~ ,
    \label{pertsolinf}
\end{align}
where a factor as $k ^{3 /2}$ is added to $\delta \varphi$ since the solutions are in coordinate phase space. Thus from the first equation one can obtain
\begin{align}
    \delta \lambda = -\frac{ a ^{-3}}{ \Omega ^{2}} \int a ^{3}( V _{\varphi} \delta \dot \varphi + V _{\varphi \varphi} \dot \varphi \delta \varphi) dt ~ .
\end{align}
%\sout{These solutions will be used as initial conditions of the perturbations in the next epoch.} 

Making use of the adiabatic-entropy decomposition proposed in \cite{Gordon:2000hv, Lalak:2007vi, Langlois:2008mn}, one can define the adiabatic and the entropy modes of those field-perturbations as
\begin{align}
    Q _{\sigma} \equiv \frac{ \dot \varphi}{ \dot \sigma} \delta \varphi +\frac{ \dot \theta}{ \dot \sigma} \delta \theta ~ ,~ Q _{s} \equiv -\frac{ \dot \theta}{ \dot \sigma} \delta \varphi +\frac{ \dot \varphi}{ \dot \sigma} \delta \theta ~ .
    \label{Qsigma and Qs}
\end{align}
where $\dot\sigma\equiv\sqrt{\dot\varphi^2+\dot\theta^2}$. Both $Q_\sigma$ and $Q_s$ are gauge-invariant variables. We can further define the curvature and isocurvature perturbations as:
\begin{align}
    {\cal R}=\frac{H}{\dot\sigma}Q_\sigma~,~~~{\cal S}=\frac{H}{\dot\sigma}Q_s~.
    \label{R and S}
\end{align}
In the inflationary epoch where $\dot \theta \simeq 0$, one has ${\cal R} \simeq \delta \varphi$ and ${\cal S} \simeq \delta \theta$. Moreover, from the solution of $\delta \varphi$ and $\delta \theta$ in \eqref{pertsolinf}, we can straightforwardly have:
\begin{align}
    {\cal R} \sim \text{const.} \times  k ^{3 /2}~,~~~{\cal S} \simeq \frac{ H}{ 2 \pi} ~ .
    \label{R and S sol inf}
\end{align}

\section{matter-dominant epoch}
\label{MDepoch}
\subsection{background}

Along with the inflaton field $\varphi$ goes down towards the minimum value of the potential, it will finally falls into the ``valley" surrounding its minimum. In order to get a period of matter-dominated epoch, we now turn on the dependence of the potential on the other fields $\theta$ and $\lambda$, and design its ansatz form to be:
\begin{align}
    V( \varphi ,\theta ,\lambda) =\frac{ 1}{ 2} \lambda( m ^{2} \varphi ^{2} +m ^{2} \theta ^{2} -\Omega ^{2}) ~ ,
    \label{potentialmd}
\end{align}
where $m$ is the effective mass of both fields $\varphi$ and $\theta$ fields. Given that the difference in masses between these two fields does not induce any new physical processes, for simplicity we assume that they have the same mass. Since in the current work $\lambda$ is set to be less than zero, the potential is actually a concave function of $\varphi$, which is consistent with the potential form \eqref{potentialinf} in the large $\varphi$ region. Assuming 
\begin{align}
    \lambda \sim a ^{-3} ~ ,
    \label{lambdamd}
\end{align}
the equation of motion for both fields \eqref{eom of fields} turns out to be
\begin{align}
    \ddot \varphi +\frac{ V _{\varphi}}{ \lambda} =0 ~,~\ddot \theta +\frac{ V_{\theta}}{ \lambda} =0 ~ .
\end{align} 
With the potential form \eqref{potentialmd}, and the initial conditions coming from the inflationary epoch (namely $\dot\varphi\simeq -\Omega$, $\dot\theta\simeq 0$), one has the solution as
\begin{align}
    \varphi =-A \Omega \sin( mt) ~,~~~\theta =A \Omega \cos( mt) ~ ,
    \label{solfieldmd}
\end{align}
with $A$ as the amplitude of both $\varphi$ and $\theta$. 

As a consistency check, we would like to claim that such a solution is consistent with the assumption \eqref{lambdamd}: With this solution, we always have $-m ^{2} \varphi ^{2} -m ^{2} \theta ^{2} +\Omega^2 \simeq 0$, therefore $V_\lambda\simeq 0$ and condition \eqref{conditionmd} is always satisfied. Therefore, we can believe that Eq. \eqref{solfieldmd} is the right solution we look for.

\begin{figure}[h]
    \includegraphics[width=250.pt]{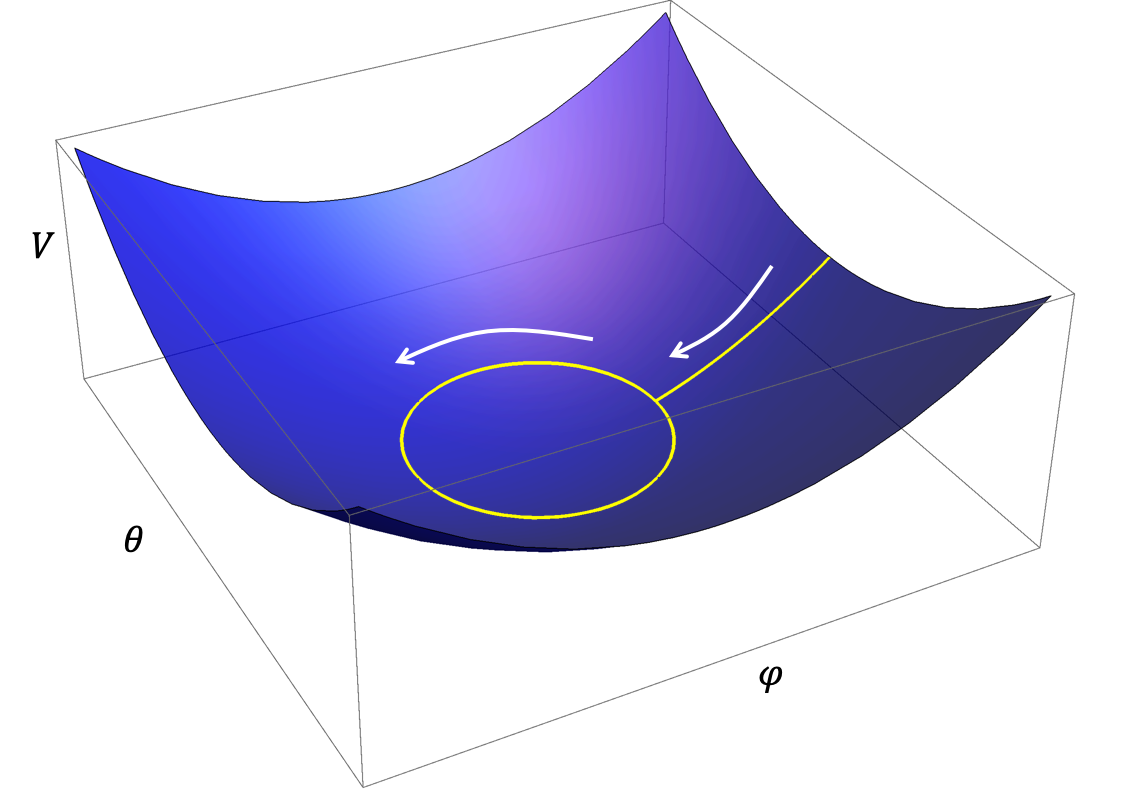}
    \caption{The sketch plot of the potential (blue) and the fields' trajectory during the matter-dominated epoch (yellow line). The white arrows indicate the direction where the fields go. Here we temporarily turn off the overall factor $\lambda$ for simplicity. }
    \label{sketch}
\end{figure}

Considering the behaviors of the fields at the inflationary epoch as well, we can summarize the overall trajectory of the system in the $\varphi-\theta$ phase space: Firstly, the system falls down the potential along the $\varphi$ direction, which causes the linear reduction of $\varphi$ field, while the $\theta$ field stays tuned with no variation either. The constant velocity of $\varphi$ gives rise to a period of inflation. Then at a certain point, the $\theta$ field is triggered and obtains similar velocity as that of $\varphi$, therefore, both the two fields start to wind around the potential and maintain at a certain altitude, due to the mimetic constraint. This gives rise to matter-dominated epoch. We show the sketch plot of the trajectory of our system in Fig. \ref{sketch}. It is difficult to present also the dependence on $\lambda$ of the potential, but from the expression \eqref{potentialmd} one can see that in the matter-dominant epoch, the potential will be decreasing as $a^{-3}$ along with the total factor $\lambda$.  

\subsection{perturbations}
Now we analyse the evolution of perturbations during matter-dominated epoch. According to Eqs. \eqref{mimetic constrain of perturbations} and \eqref{eom of perturbations of fields} with the background solution \eqref{lambdamd} and \eqref{solfieldmd}, one gets the perturbation equations for this epoch to be
\begin{align}
   \delta \ddot \varphi +\left( \frac{ k ^{2}}{ a ^{2}} +m ^{2} \right) \delta \varphi +\left( 3 H \delta \lambda +\delta \dot \lambda \right) \frac{ \dot{ \varphi}}{ \lambda} &=&0 ~ , \nonumber \\
   \delta \ddot \theta +\left( \frac{ k ^{2}}{ a ^{2}} +m ^{2} \right) \delta \theta +\left( 3 H \delta \lambda +\delta \dot \lambda \right) \frac{ \dot{ \theta}}{ \lambda} &=&0 ~ ,
   \label{eom of deltavarphi and deltatheta md}
\end{align}
while the constraint equation is 
\begin{align}
    \dot{ \theta} \delta \dot{ \theta} +\dot{ \varphi} \delta \dot{ \varphi} -m ^{2} \varphi \delta \varphi -m ^{2} \theta \delta \theta =0 ~ .
    \label{constrain of perturbations md}
\end{align}

Thanks to the triangle functional forms of the background solution of $\varphi$ and $\theta$, some terms in Eqs. \eqref{eom of deltavarphi and deltatheta md} get cancelled so that the dependence of $\delta \lambda$ in these two equations has a unitary form and can be removed using the constraint equation. Taking time derivative to the constraint equation \eqref{constrain of perturbations md} one gets
\begin{align}
    \dot{ \theta} \delta \ddot{ \theta} +\dot{ \varphi} \delta \ddot{ \varphi} +(\ddot{ \theta} -m ^{2} \theta) \delta \dot{ \theta} +(\ddot{ \varphi} -m ^{2} \varphi) \delta \dot{ \varphi} \nonumber \\
    -m ^{2} \dot \varphi \delta \varphi -m ^{2} \dot \theta \delta \theta =0 ~ .
    \label{constrain of perturbations md2}
\end{align}
Rearrange Eqs. \eqref{eom of deltavarphi and deltatheta md} and \eqref{constrain of perturbations md2}, one finally gets:
%\begin{widetext}
%\begin{eqnarray}
%    \delta\ddot\phi-\frac{2\phi\dot\phi}{A^2\Omega^2}\delta\dot\phi-2\frac{\theta\dot\phi}{A^2\Omega^2}\delta\dot\theta+\left(\frac{k^2}{a^2}+m^2\right)\left[\left(1-\frac{(\dot\phi)^2}{A^2m^2\Omega^2}\right)\delta\phi
%    -\frac{\dot\phi\dot\theta}{A^2m^2\Omega^2}\delta\theta\right]&=&0~,\\
%      \delta\ddot\theta-\frac{2\theta\dot\theta}{A^2\Omega^2}\delta\dot\theta-2\frac{\phi\dot\theta}{A^2\Omega^2}\delta\dot\phi+\left(\frac{k^2}{a^2}+m^2\right)\left[\left(1-\frac{(\dot\theta)^2}{A^2m^2\Omega^2}\right)\delta\theta 
%    -\frac{\dot\theta\dot\phi}{A^2m^2\Omega^2}\delta\phi\right]&=&0~,\\
%   \label{eom of perturbation of fields md} \delta\dot\lambda+3H\delta\lambda+\lambda\left(\frac{k^2}{a^2}+m^2\right)\frac{\dot\theta\delta\theta+\dot\varphi\delta\varphi}{A^2m^2\Omega^2}&=&0~.
%   \label{eom of delta lambda md}
%\end{eqnarray}
%\end{widetext}
\begin{widetext}
\begin{eqnarray}
    \delta \ddot \varphi +\left( \frac{ k ^{2}}{ a ^{2}} +m ^{2} \right) \delta \varphi -\frac{ \dot{ \varphi}}{ A ^{2} m^{2} \Omega ^{2}} \left[ \left( \frac{ k ^{2}}{ a ^{2}} +2 m ^{2} \right)( \dot{ \theta} \delta \theta +\dot{ \varphi} \delta \varphi) +2 m ^{2}( \theta \delta \dot{ \theta} +\varphi \delta \dot{ \varphi}) \right] &=&0 ~ ,\\
    \delta \ddot \theta + \left( \frac{ k ^{2}}{ a ^{2}} +m ^{2} \right) \delta \theta -\frac{ \dot{ \theta}}{ A ^{2} m ^{2} \Omega ^{2}} \left[ \left( \frac{ k ^{2}}{ a ^{2}} +2 m ^{2} \right)( \dot{ \theta} \delta \theta +\dot{ \varphi} \delta \varphi) +2 m ^{2}( \theta \delta \dot{ \theta} +\varphi \delta \dot{ \varphi}) \right] &=&0 ~ ,\\
   \label{eom of perturbation of fields md}
   \delta \dot \lambda +3 H \delta \lambda +\frac{ \lambda}{ A ^{2} m ^{2} \Omega ^{2}} \left[ \left( \frac{ k ^{2}}{ a ^{2}} +2 m ^{2} \right)( \dot{ \theta} \delta \theta +\dot{ \varphi} \delta \varphi) +2 m ^{2}( \theta \delta \dot{ \theta} +\varphi \delta \dot{ \varphi}) \right] &=&0 ~ .
   \label{eom of delta lambda md}
\end{eqnarray}
\end{widetext}
From the above equations one can see that, in this epoch the equations of field-perturbations actually don't depend on $\delta \lambda$, which simplifies our calculations.

%{\green from eq(32), in MD era we have the equations of field-perturbations $\delta\ddot\phi^a+(\frac{k^2}{a^2}+m^2)\delta\phi^a+(3H\delta\lambda+\delta\dot\lambda)\frac{\dot{\phi}}{\lambda}=0$ and the constrain(33) $\dot{\theta}\delta\dot{\theta}+\dot{\phi}\delta\dot{\phi}-m^2\phi\delta\phi-m^2\theta\delta\theta=0$ (here for simply discussing, we neglect the metric perturbation which has no effect on take $\delta\lambda$ apart from the equations of field-perturbations).

%Then take the derivative of the constrain with respect to time, we have $\dot{\theta}\delta\ddot{\theta}+\dot{\phi}\delta\ddot{\phi}+F_1(\phi^a,\dot{\phi}^a,\delta\phi^a,\delta\dot{\phi}^a)=0$

%Replace $\delta\ddot{\theta}$ and $\delta\ddot{\phi}$ in constrain with the equations of field-perturbations, we have $-\frac{\dot{\theta}^2+\dot{\phi}^2}{\lambda}(3H\delta\lambda+\delta\dot\lambda)+F_2(\phi^I,\dot{\phi}^I,\delta\phi^I,\delta\dot{\phi}^I)=0$. Notice that $\dot{\theta}^2+\dot{\phi}^2=m^2 A^2$, so we get the equation of $\delta\lambda$:  $\delta\dot\lambda+3H\delta\lambda+\lambda F_3(\phi^I,\dot{\phi}^I,\delta\phi^I,\delta\dot{\phi}^I)=0$ and rewrite  the equations of field-perturbations $\delta\ddot\phi^a+(\frac{k^2}{a^2}+m^2)\delta\phi^a-F_3(\phi^I,\dot{\phi}^I,\delta\phi^I,\delta\dot{\phi}^I)\frac{\dot{\phi}}{\lambda}=0$}
Using the definition of the adiabatic and the entropy perturbation modes given in \eqref{Qsigma and Qs}, one has
%\begin{eqnarray}
%   Q_\sigma^{\prime\prime}+\frac{\kappa^2}{a^2}Q_\sigma+2Q_s^\prime&=&0~,\\
%   \label{eq of Qsigma}
%    Q_s^{\prime\prime}+Q_s &=&0~,    
%\end{eqnarray} 
\begin{eqnarray}
    Q _{\sigma} ^{\prime \prime} &=& 0 ~ ,\\
    \label{eq of Qsigma}
    Q _{s} ^{\prime \prime} +\frac{ \kappa ^{2}}{ a ^{2}} Q _{s} +2 Q _{\sigma} ^{\prime} &=& 0 ~,    
\end{eqnarray}
where $\prime$ denotes derivative with respect to the normalized time $x \equiv m t$, and $\kappa \equiv k /m$ is also the normalized wave-number. In matter-dominated epoch, the scale factor and the Hubble parameter are $a( x) =a _{0} x ^{2 /3}$ and ${\cal H}( x) =2 m /(3 x)$ respectively. 
%From the equations we can see that, the entropy mode $Q_s$ acts as a source of the adiabatic mode $Q_\sigma$. 
Furthermore, from the definition of the curvature and isocurvature perturbations given in \eqref{R and S}, one has equations as
%\begin{eqnarray}
%    {\cal R}^{\prime\prime}+3{\cal H}{\cal R}^\prime+\frac{\kappa^2}{a^2}{\cal R}+2{\cal S}^\prime+3{\cal H}{\cal S}&=&0~,\\
%    {\cal S}^{\prime\prime}+3{\cal H}{\cal S}^\prime+{\cal S}&=&0~.
%\end{eqnarray}
\begin{eqnarray}
    {\cal R} ^{\prime \prime} +3 {\cal H} {\cal R} ^{\prime} &=&0 ~ ,\\
    {\cal S} ^{\prime \prime} +3 {\cal H} {\cal S} ^{\prime} +\frac{ \kappa ^{2}}{ a ^{2}} {\cal S} +2 {\cal R} ^{\prime} +3 {\cal H} {\cal R} &=&0 ~ .    
\end{eqnarray}
Interestingly, we can see that the curvature perturbation ${\cal R}$ satisfies a homogeneous equation of motion which is seemingly not dependent on $k$. However, it is not true since it is also influenced by the initial conditions, as we will see later.
%Again we can see, the isocurvature perturbation ${\cal S}$ will act as a source of the curvature perturbation ${\cal R}$, and the latter will be generated from the former. 

The above equations gives the solutions
%\begin{eqnarray}
%    Q_s&=&C_1 e^{ix}+C_2 e^{-ix}~,\\
%    Q_\sigma&=&C_3 L_1(x)+C_4 L_2(x) \nonumber\\
%&&+L_2(x)\int_0^{x}\frac{L_1(x^\prime)(-2Q_s^\prime(x^\prime))}%{W(x^\prime,L_1,L_2)}dx^\prime \nonumber\\
%&&-L_1(x)\int_0^{x}\frac{L_2(x^\prime)(-2Q_s^\prime(x^\prime))}{W(x^\prime,L_1,L_2)}dx^\prime~,
%\end{eqnarray}
\begin{eqnarray}
    Q _{\sigma} &=& C _{1} +C _{2}( x -x _{0}) , \nonumber \\
    Q _{s} &=& C _{3} L _{1}( x) +C _{4} L _{2}( x) -2 C _{2} L _{3}( x) ~ ,
    \label{Qsigma and Qs md}
\end{eqnarray}
where 
%\begin{eqnarray}
%    L_1(x)&=&-3px^{1/3}\cos(3px^{1/3})+\sin(3px^{1/3})~\text{ and }\\
%    L_2(x)&=&\cos(3px^{1/3})+3px^{1/3}\sin(3px^{1/3})
%\end{eqnarray}
\begin{eqnarray} 
    L _{1}( x) &=& -3 p x ^{1 /3} \cos( 3 p x ^{ 1 /3}) +\sin( 3 p x ^{1 /3}) , \nonumber \\
    L _{2}( x) &=& \cos( 3 p x ^{1 /3}) +3 p x ^{1 /3} \sin( 3 p x ^{1 /3}) , \text{ and}\nonumber \\
    L _{3}( x) &=& -\frac{ 8}{ 81 p ^{6}} -\frac{ 4 x ^{2 /3}}{ p ^{4}} +\frac{ x ^{4 /3}}{ p ^{2}} ,
\end{eqnarray}
are the general solutions for the homogeneous part of Eq. \eqref{eq of Qsigma}, %$W(x,L_1, L_2)=-9p^3$ is the Wronski determinant for the general solutions $L_1$ and $L_2$, 
and $p \equiv \kappa /a _{0}$. Here we set the initial time of this epoch to be $x =x_0$. When the perturbations reentered the horizon in matter-dominated epoch, one has $k _{\ast} =a( x _{\ast}) H( x _{\ast})$, which gives rise to $x _{\ast} ^{1 /3} p =2 /3$. 

In order to get the exact perturbations during the matter-dominated epoch, we should use the continuity conditions to match the solutions between \eqref{Qsigma and Qs md} and those of the inflationary epoch, namely Eq. \eqref{R and S sol inf}. Here we consider two kinds of continuity conditions. One is to consider the continuity of the curvature/isocurvature perturbations themselves, namely ${\cal R}$ and ${\cal S}$. In this case, one obtains the initial conditions for the matter-dominated epoch as
%{\green To get the perturbations evolution during the Matter-Dominated  , we consider to use the continuous conditions of curvature perturbation and isocurvature perturbation at the end of inflation , which provide the initial conditions for the MD epoch.
\begin{eqnarray}
    \mathcal{ R} _{0} &=& \delta \varphi _{0} \equiv \beta \frac{ H}{ 2 \pi} p ^{3 /2} ~ , \quad \mathcal{ R} _{0} ^{\prime} =0 ~ , \nonumber \\
    \mathcal{ S} _{0} &=& \delta \theta _{0} =\frac{ H}{ 2 \pi} ~ , \quad \mathcal{ S} _{0} ^{\prime} =0 ~ ,
\end{eqnarray}
where $\beta$ is a constant scaling the amplitude of $\delta \varphi$. Substituting the above conditions into Eq. \eqref{Qsigma and Qs md} one can get the exact solution as
\begin{eqnarray}
    \mathcal{ R} &=& \frac{ 2}{ 3 x}( C _{1} +C _{2} x)\delta\theta _{0} ~ , \nonumber \\
    \mathcal{ S} &=& \frac{ 2}{ 3 x}[ C _{3} L _{1}( x) +C _{4} L _{2}( x) -2 C _{2} L _{3}( x)]\delta\theta _{0} ~ ,
\end{eqnarray}
where 
\begin{eqnarray}
    C_{1} &=& 0 , \quad C _{2} =\frac{ 3}{ 2} \beta p ^{3 /2} ~ , \nonumber \\
    \frac{ C _{3}}{ x} &=& \frac{ 1}{ \gamma ^{3}} \left[ \left( \frac{ 1}{ 6} -\frac{ 1}{ 2} \gamma _{0} ^{2} \right) \cos( 3 \gamma _{0}) +\frac{ 1}{ 2} \gamma _{0} \sin( 3 \gamma _{0}) \right] \nonumber \\
    &+& \frac{ \beta p ^{1 /2}}{ \gamma ^{3}} \bigg[ \left( \frac{ 16}{ 81} -\frac{ 8}{ 9} \gamma _{0} ^{2} \right) \sin( 3 \gamma _{0}) +\bigg( \frac{ 2}{ 3} \gamma _{0} ^{3} \nonumber \\
    &-& \frac{ 16}{ 27} \gamma _{0} \bigg) \cos( 3 \gamma _{0}) \bigg] \propto k ^{0} +\mathcal{ O}( k ^{3 /2}) ~ , \nonumber \\
    \frac{ C _{4}}{ x} &=& \frac{ 1}{ \gamma ^{3}} \left[ \frac{ 1}{ 2} \gamma _{0} \cos{ \left( 3 \gamma _{0} \right)} -\left( \frac{ 1}{ 6} -\frac{ 1}{ 2} \gamma _{0} ^{2} \right) \sin{ \left( 3 \gamma _{0} \right)} \right] \nonumber \\
    &+& \frac{ \beta p^{ 3 /2}}{ \gamma ^{3}} \bigg[ \left( \frac{ 16}{ 81} -\frac{ 8}{ 9} \gamma _{0} ^{2} \right) \cos{ \left( 3 \gamma _{0} \right)} \nonumber \\
    &-& \left( \frac{ 2}{ 3} \gamma _{0} ^{3} -\frac{ 16}{ 27} \gamma _{0} \right) \sin{ \left( 3 \gamma _{0} \right)} \bigg] \propto k ^{0} +\mathcal{ O} \left( k ^{3/2} \right) \nonumber \\   
    \frac{ C _{2}}{ x} L _{3} &=& \frac{ \beta}{ p ^{3 /2}} \left( \gamma -\frac{ 4}{ 9} \gamma ^{-1} -\frac{ 8}{ 81} \gamma ^{-3} \right) \propto k ^{-3 /2} ~ ,
\end{eqnarray}
where $\gamma \equiv x ^{1 /3} p \in( 0 ,2 /3]$, and $\gamma _{0} =x _{0} ^{1 /3} p \propto k \ll 1$. For the curvature perturbation, it obviously follows $\mathcal{R}\propto k^{3/2}$ with a spectral index $n_s=4$, while the isocurvature perturbation ${\cal S}$ can obtain a scale-invariant spectrum by requiring $\beta \ll 1$. %However, this would result in the isocurvature perturbation surpassing the curvature one. In fact, such a scenario is permissible since distinguishing between these two perturbations is challenging. Thus, a dominant scale-invariant perturbation aligns with observations, whether it manifests as a curvature or isocurvature perturbation. 
Furthermore, on a larger scale, due to the dominance of the $C _{3} L _{3} /x$ term, the isocurvature perturbation will have a red-tilted spectrum, characterized by $n _{s} =-2$. %Consequently, by adjusting parameters appropriately, we can establish a period of scale invariance within our area of interest.
\begin{figure}[h]
%    \subfigure[]{
    \includegraphics[width=250.pt]{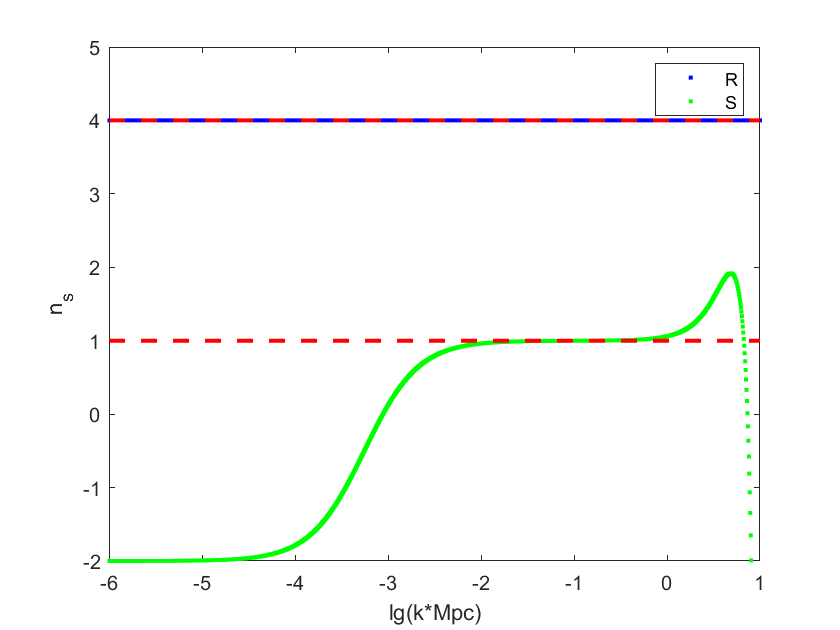}
    \label{amplitude}
%    }
%    \subfigure[]{
%    \includegraphics[width=200.pt]{p4.png}
%    \label{index}
%    }  
    \caption{Plots of the spectral indices of the curvature and the isocurvature perturbation power spectrum, $P _{\cal R}$ and $P _{\cal S}$. The parameters and initial conditions are chosen as $m =10 ^{-5} M _{\text{P}}$, $A =10 ^{5} M _{\text{P}} ^{-1}$, 
    $\delta \theta( 0) =H /2 \pi$, $\beta =10 ^{-2}$, $x _{0} =10 ^{-8}$.}
\end{figure}

Another continuous condition is to consider the continuity of the adiabatic and the entropy perturbation modes $Q _{\sigma}$ and $Q _{s}$, which can give rise to another set of initial conditions as
\begin{eqnarray} 
    Q _{s}( x =x _{0}) &=& -\delta \theta _{0} ~,~Q _{s} ^{\prime}( x =x _{0}) =\delta \phi _{0} =\beta p^{3 /2}\delta \theta _{0} ~, \nonumber \\
    Q _{\sigma}( x =x _{0}) &=& -\delta \phi _{0} ~,~ Q _{\sigma} ^{\prime}( x =x _{0}) = -\delta \theta _{0}.
\end{eqnarray}
Following a similar way, we can derive the expressions for curvature and isocurvature perturbations:
\begin{eqnarray}
    \mathcal{ R} &=& -\delta \theta _{0}[ 1 +(\beta p ^{3 /2} -x _{0}) p ^{3} \gamma ^{-3}] ~ , \nonumber \\
    \mathcal{ S} &=& -\delta \theta _{0}( S _{1} +\beta p ^{3 /2} S _{2}) ~ ,
\end{eqnarray}
where
\begin{eqnarray}
    S _{1} &=& \frac{ 1}{ p ^{3}} \left( \frac{ 16}{ 81} \gamma ^{-3} -\frac{ 16}{ 27} \gamma ^{-2} \sin( 3 \gamma -3 \gamma _{0}) +\frac{ 8}{ 9} \gamma ^{-1} -2 \gamma \right) \nonumber \\
    &+& \frac{ p ^{2}}{ x _{0} ^{1 /3}} \left[ \frac{ \cos( 3 \gamma -3 \gamma _{0})}{ \gamma ^{2}} -\frac{ \sin( 3 \gamma -3 \gamma _{0})}{ 3 \gamma ^{3}} \right] \nonumber \\
    &+& \frac{ x _{0} ^{1 /3}}{ p ^{2} \gamma} \bigg[ \left( \frac{ 8 \gamma _{0}}{ 9 \gamma ^{2}} +\frac{ 2 \gamma _{0} ^{2}}{ \gamma} -\frac{ 16}{ 9 \gamma} \right) \cos( 3 \gamma -3 \gamma _{0}) \nonumber \\
    &+& \left( \frac{ 16}{ 27 \gamma ^{2}} +\frac{ 8 \gamma _{0}}{ 3 \gamma} -\frac{ 2}{ 3 \gamma} \right) \sin( 3 \gamma -3 \gamma _{0}) \bigg] ~ , \nonumber \\
    S _{2} &=& \frac{ 1}{ 9 \gamma ^{3}}[ 3( \gamma -\gamma _{0}) \cos( 3 \gamma -3 \gamma _{0}) \nonumber \\
    &-& (1 +3 \gamma _{0} \gamma) \sin( 3 \gamma -3 \gamma _{0})] ~ .
\end{eqnarray}
From the above we can see that, under the condition $\beta p ^{3 /2} \ll x _{0} =\gamma _{0} p ^{-3} \ll 1$, we can obtain a scale-invariant curvature perturbation $\mathcal{ R} \sim k ^{0} +\mathcal{ O}( k ^{9 /2})$, while the isocurvature perturbation is red-tilted, $\mathcal{ S} \sim k ^{-3}$. Moreover, on small scales both curvature and isocurvature perturbations will have a blue-tilt, which can be responsible for primordial black hole generation.  %However, we notice a persistent issue with dominant non-scale-invariant isocurvature perturbations, regardless of the parameters selection. {\green This arises due to the absence of terms in the isocurvature perturbation lacking dependence on $k$,} which fails to align with observed results.

\begin{figure}[h]
%    \subfigure[]{
    \includegraphics[width=250.pt]{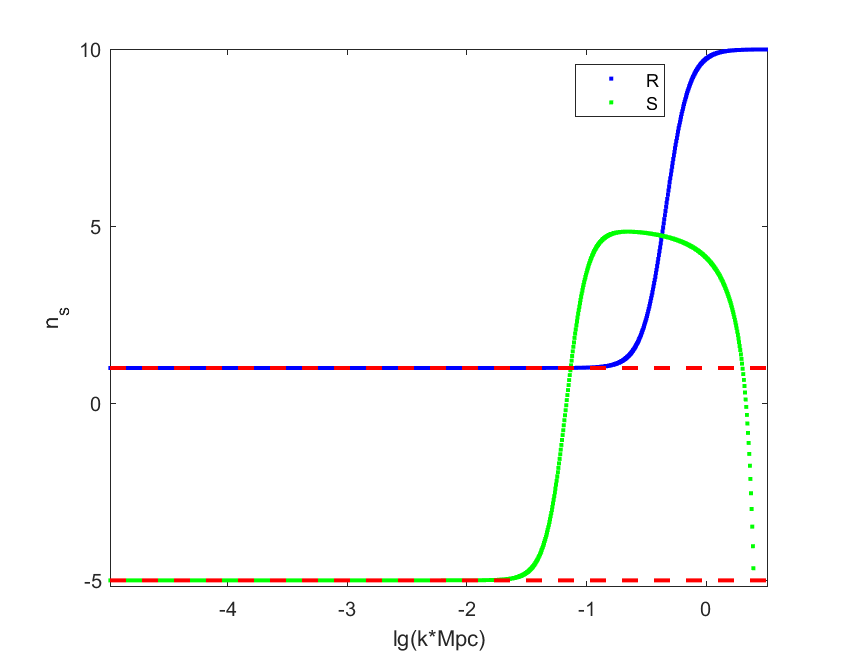}
    \label{amplitude}
%    }
%    \subfigure[]{
%    \includegraphics[width=200.pt]{2.png}
%    \label{index}
%    }  
    \caption{Plots of the spectral indices of curvature and isocurvature perturbations power spectra, $P _{\cal R}$ and $P _{\cal S}$. The parameters and initial conditions are chosen as $m =10 ^{-2} M _{\text{p}}$, $A =10 ^{2} M _{\text{p}} ^{-1}$, $\delta \theta( 0) =H /2 \pi$, $\beta =10 ^{-2}$, $x _{0} =10 ^{-5}$.}
\end{figure}
\section{conclusions}
\label{conclusions}

Mimetic gravity is an interesting gravity theory that can mimic the matter section, by adding a constraint to the auxiliary scalar field in a pure gravity (or gravity plus the auxiliary scalar) framework. When the auxiliary scalar acts as an inflaton field, this theory can connect the inflationary and matter-dominated epochs. However, in both ordinary one-field and two-field mimetic inflation models, the perturbations will have problems. Especially in the two-field model, the kinetic term of the adiabatic mode will not appear in the action because of the constraint, making the adiabatic perturbations seemingly not propagate. In our previous paper, we proposed a curvaton mechanism to make the adiabatic perturbation generated from the entropy perturbation, where we take the parameter $\lambda$ to be a pure multiplier and thus neglect its perturbation, $\delta \lambda$. In order to complete our analysis, in the current paper, we treat $\lambda$ as a non-dynamical field and thus $\delta \lambda$ is also involved. 

For both inflationary and matter-dominated epochs, we analysed the evolution of the mimetic fields in background and perturbation level. In inflationary epoch, with the potential designed to be nearly flat, we found that due to the constraint imposes to the fields, the inflaton field rolls down along the potential with a constant velocity, while the curvaton field remains still. This background solution is the same as that of nontrivial field metric in our previous paper, and this means that the inflationary trajectory is not quite sensitive to the factors of the field metric. Moreover, the constraint at perturbation level induces the inflaton perturbation to be a constant, while the curvaton perturbation has the usual solution of $H /2 \pi$. These results will act as the initial conditions of the subsequent matter-dominated epoch.

In the matter-dominated epoch, we design the potential such that it has quadratic shape for both the inflaton and curvaton field, therefore, both two fields can have oscillating behaviors. This solution satisfies the constraint equation, and can make the energy density of the universe (which is mainly given by $\lambda$) to be matter-like. This gives the whole trajectory of the model which, after rolling down towards the potential, will wind around it at some height near the minimum. Moreover, in this epoch the perturbations of the two fields can be independent on $\delta\lambda$. We transfer the field perturbations into the adiabatic/isocurvature perturbations and solve the equations of motion to get their evolution. We find that the $k$-dependence of the curvature and isocurvature perturbations depends on the continuity conditions between inflationary and matter-dominated epoch. For the ${\cal R} /{\cal S}$-continuity conditions, one can get a nearly scale invariant isocurvature perturbation and a strong blue-tilted curvature perturbation on small scales, but on large scales the isocurvature perturbation will be red-tilted. For the $Q$-continuity conditions, one can get a scale-invariant curvature perturbation and a red-tilted isocurvature perturbation, and blue-tilted curvature perturbation on small scales. This will be responsible for primordial black hole generations, which will be discussed in a separate future work.

Our results show that whether $\lambda$ is a mathematical multiplier or a non-dynamical field, the curvaton mechanism will generate scale-invariant perturbations and power spectra approved by cosmic microwave background observations. This means that the mimetic curvaton model can basically work as the early universe model, driving both inflationary and matter-dominated epochs. The next issues we will investigate include extensive applications to various aspects in cosmology, such as primordial black hole generation, or some non-singular model buildings like bounce inflation model. We will deal with these subjects in our future works.

\begin{acknowledgments}
We acknowledge Lei-Hua Liu, Seyed Ali Hosseini Mansoori and Alexander Vikman for helpful discussion at the earlier stage of this work. This work is supported by the National Key Research and Development Program of China (Grant No. 2021YFC2203100).

\end{acknowledgments}

\bibliographystyle{apsrev4-1}
\bibliography{bibfile.bib}

\end{document}